\documentclass[11pt]{JHEP3}
\usepackage{graphicx}
\usepackage{amsmath, amssymb}

\newcommand{\be}{\begin{eqnarray}}
\newcommand{\ee}{\end{eqnarray}}
\newcommand{\nn}{\nonumber}
\newcommand{\bn}{\begin{enumerate}}
\newcommand{\en}{\end{enumerate}}

\def\IR{\mathbb{R}}

\def\CE{{\cal E}}
\def\CF{{\cal F}}

\def\CL{{\cal L}}

\def\CN{{\cal N}}
\def\CO{{\cal O}}

\def\a{\alpha}

\def\d{\delta}
\def\e{\epsilon}

\def\th{\theta}

\def\l{\lambda}
\def\m{\mu}
\def\n{\nu}

\def\s{\sigma}

\def\vph{\varphi}
\def\w{\omega}

\def\D{\Delta}

\def\O{\Omega}

\def\half{\frac{1}{2}}
\def\thalf{{\textstyle \frac{1}{2}}}
\def\imp{\Longrightarrow}
\def\goto{\rightarrow}

\def\p{\partial}

\def\Tr{{\rm Tr}}

\def\det{{\rm det}}

\def\fft#1#2{{#1 \over #2}}

\title{Near Horizon Analysis of Extremal AdS$_\mathbf{5}$
Black Holes}

\author{Jaehyung Choi$^1$, Sungjay Lee$^2$, Sangmin Lee$^2$\\
\\ 
$^1$ 
Department of Physics and Astronomy, 
SUNY,
Stony Brook, NY 11794-3800, USA
\\
$^2$ 
School of Physics and Astronomy, Seoul National University, 
Seoul 151-747, Korea
}

\abstract{
We study the near horizon geometry of extremal black holes in five dimensional gauged supergravity using Sen's entropy function formalism.
Special attention is paid to the large black hole limit where 
the near horizon solution exhibits a universal dependence 
on the rotation. 
The physical properties of the large black hole solution
are shown to agree with predictions from 
fluid mechanical description of the dual conformal field theory.  
}


\preprint{SNUST-080202}

\begin{document}


\section{Introduction}

The AdS/CFT correspondence \cite{adscft} is 
a concrete realization of 
the holographic principle \cite{holo1,holo2,holo3} 
in the sense that the CFT defined 
on the boundary of the AdS space is believed to 
capture the full dynamics of string theory 
in the bulk. One may say that the boundary 
of the AdS serves as the ``holographic screen."

Given that the holographic principle, including the notion of 
the holographic screen, has its origin in black hole thermodynamics,  
the recurring appearance of black holes in 
the development of the AdS/CFT correspondence 
should come as no surprise. 
For example, recent applications of the AdS/CFT methods  
to the hydrodynamic regime of strongly coupled field theories
\cite{hy1,hy2,hy3,hy4,nat1,hy5}  
bear strong resemblance to the membrane paradigm of 
black hole physics \cite{mem1,mem2,mem3}.

When the AdS space contain a black hole inside, 
one encounters an interesting situation 
where the two holographic screens $-$ the black hole horizon 
and the AdS boundary $-$ exist at the same time. 
It appears that questions on black hole dynamics can 
be addressed from the two ``dual'' points of view. 
For instance, while traditional approaches to understand the 
black hole entropy 
have often considered degrees of freedom living on the horizon, 
but AdS/CFT suggests that the dual CFT on the AdS boundary 
should give a microscopic explanation of the entropy. 

If both the traditional picture and the AdS/CFT picture make sense, 
it is conceivable that there exists some mapping which 
relate physical quantities on the two holographic screens 
as well as the equations governing them. 
Exploring the possibility of such a mapping was a key motivation 
which initiated this work. 

Recently, in \cite{minwa}, the description of the CFT in terms 
of fluid dynamics was used to make striking predictions on 
the thermodynamics of black holes in AdS. 
Fluid dynamics is valid when the fluctuation of the CFT is macroscopic. 
This translates to the condition that the black hole should 
be ``large'' in a technical sense we will review in section 2. 
Among other things, the result of \cite{minwa} shows that 
large black holes exhibit universal dependence on the 
rotation parameters.

The main goal of this paper is to study the properties 
of large black holes from the opposite side, namely, 
the black hole horizon, bearing in mind the original motivation mentioned above.
For simplicity, we focus on the extremal limit where 
the entropy function formalism \cite{sen1,sen2,sen5} 
enables us to extract physical informations without dealing with 
all the field equations. Although most of our analysis 
does not rely on supersymmetry or number of dimensions, 
we focus on black holes in 
gauged supergravity in five dimension 
because AdS$_5$/CFT$_4$ with supersymmetry 
offer more examples of dual pairs than any other cases, 
of which both sides of the duality have explicitly 
known Lagrangian descriptions.

We derive the near horizon equations of motion and 
the entropy function for general extremal black holes in AdS$_5$. 
Then we specialize to the large black hole limit 
and verify that, to the leading order, the near horizon equations 
admit a universal solution with a factorized dependence 
on the rotation parameters. 
We then make a detailed comparison with the predictions from fluid dynamics 
of \cite{minwa} and find perfect agreement. 
In addition, we show how the near horizon equations  
determine the charge dependence of the thermodynamic potentials 
(at zero temperature), which cannot be inferred from the 
analysis of fluid dynamics alone.

This paper is organized as follows. 
Section 2 covers some preliminary materials. 
We first establish our notations on gauged supergravity. 
Then we briefly review the predictions from fluid dynamics \cite{minwa}
on the properties of large black holes in AdS.  
We also illustrate salient features of the 
large black hole limit using a simple example in the minimal supergravity. 
Section 3 and 4 present the main computations of this paper.  
In section 5, we elaborate on the comparison between our results 
and the results from fluid dynamics, comment on the 
implications of our results on the thermodynamic property of the CFT, 
and conclude with some future directions.


\section{Preliminaries}

\subsection{Gauged supergravity in five dimension}

We will work with $D=5$, $\CN=1$ gauged supergravity theories 
with $n$ abelian vector fields. 
The bosonic part of the Lagrangian is
\be
\label{lag}
(16 \pi G)\CL = \star \left( R  - 2 V \right)
- \thalf g_{ij} d\vph^i \wedge \star d\vph^j 
 - \thalf Q_{IJ} F^I\wedge \star F^J 
- \textstyle{\frac{1}{6}} C_{IJK}F^I\wedge F^J \wedge A^K.
\ee 
The $(n-1)$ scalars, $\vph^i$, parameterize a hypersurface 
in $\IR^n$,
\be
\label{hs}
\frac{1}{6} C_{IJK}X^IX^JX^K = 1. 
\ee
The real constant coefficients $C_{IJK}$ define the ``real special geometry,''
\footnote{
We follow the conventions of \cite{gut3} except that 
$Q_{IJ}^{\text{here}} = 2 Q_{IJ}^{\text{there}}$ and 
$X_I^{\text{here}} = 3X_I^{\text{there}}$.}
\be
X_I \equiv \half C_{IJK} X^JX^K, 
\;\;\;\;\;
Q_{IJ} \equiv X_IX_J - C_{IJK}X^K, 
\;\;\;\;\;
g_{ij} \equiv Q_{IJ} \partial_i X^I \partial_j X^J. 
\ee
The scalar potential in (\ref{lag}) is specified by some constants 
$\bar{X}_I$ :
\be
V = Q^{IJ}\bar{X}_I\bar{X}_J- (X^I\bar{X}_I)^2 .
\ee
The supersymmetric extremum of the potential is located at $X_I|_* = \bar{X}_I$, 
where $V|_* = -6$. This amounts to setting the AdS radius to be unity: 
$R_{\m\n} = -4  g_{\m\n}$. 

All supergravity theories of the form (\ref{lag}) contain 
the minimal supergravity as a closed subsector, as one can see by setting 
\be
\label{mini}
X_I = \bar{X}_I, \;\;\;\;\; A^I = \bar{X}^I A . 
\ee
Classically, the overall normalization of $C_{IJK}$ is a matter of 
convention, since the Lagrangian (\ref{lag}) is invariant under 
\be
C_{IJK} \goto \l^3 C_{IJK}, 
\;\;\; 
A^I \goto \l^{-1} A^I, 
\;\;\; 
X^I \goto \l^{-1} X^I.
\ee 

There are many examples of gauged supergravity theories 
of which the dual CFT is known explicitly.  
Upon truncation to the massless abelian sector, 
the famous AdS$_5\times S^5$ string theory 
leads to the $U(1)^3$ theory with $C_{123}=1$ and 
all other components of $C_{IJK}$ vanishing. 
Orbifolds of $S^5$ corresponds to quiver gauge theories. 
Another very large class of $\CN=1$ CFTs 
arise from D3-branes probing toric Calabi-Yau cones 
and are efficiently described by the brane tiling model \cite{dimer1,dimer2,dimer3}. 
The coefficients $C_{IJK}$ of the corresponding supergravity theories 
are given by the area of the triangles in the toric diagram \cite{tachi,me1}. 

\subsection{Predictions from fluid dynamics}

According to the AdS/CFT correspondence, 
a black hole in a global AdS space corresponds 
to a thermal ensemble of the states in the dual CFT 
with the same 
quantum numbers as the black hole.  
In the AdS$_5$/CFT$_4$ case under discussion, 
the relevant quantum numbers are the energy $E$, the charges $Q_I$ and 
the two angular momenta $J_a$. 

At sufficiently high temperature and/or density, 
the CFT is expected to admit an effective description 
in terms of fluid mechanics. 
In \cite{minwa}, this expectation was combined with known 
properties of static black holes to make predictions on 
rotating black holes, which were then verified in all known 
rotating black hole solutions in the literature. 
We briefly summarize the result of \cite{minwa} here.

In the static case, conformal invariance and extensivity dictates that the
grand canonical partition function of the fluid take the form
\begin{equation} \label{pffom:eq}
\ln Z_{\rm gc} =  V T^3  h(\mu /T)\,,
\end{equation}
where $\mu^I$ are the chemical potentials conjugate to the 
charges $Q_I$. 
$V$ and $T$ represent the volume and the overall temperature of the fluid.
 
It is not known how to compute the function $h(\mu/T)$ 
or the equation of state of the static fluid 
directly from the CFT. 
But, if a charged static black hole solution is known, 
they can be read off using AdS/CFT. 
The equation of state of the static fluid 
is taken as an input into the relativistic Navier-Stokes equations that govern the dynamics of the conformal fluid in general.

A key observation of \cite{minwa} is that 
there exists a unique family of ideal fluid solutions 
to the Navier-Stokes equation 
in one to one correspondence with rotating black holes, 
which are simple enough to be written down explicitly. 
In the five dimensional case, the solution can be summarized 
by the grand canonical partition function for the rotating fluid, 
\begin{equation}\label{finalx}
 \ln Z_{\rm gc} =
 \ln \Tr \exp \left[ -\frac{(E - \mu^I Q_I - \O^a  J_a)}{T} \right]
 = \frac{V T^3 h(\mu/T)}
       {(1-\O_1^2)(1-\O_2^2)}\,,
\end{equation}
where $E$ and $\O^a$ represent the energy 
and the angular velocities of the fluid respectively. 
Note that the solution is universal in the sense that 
the rotation dependence factorizes and is independent of 
the function $h(\mu/T)$. 
All the physical observables, such as the energy, entropy, charges 
and angular momenta can be obtained by differentiating 
(\ref{finalx}) by the conjugate variables.

Fluid dynamics becomes a good description of the CFT if and only if 
the ``mean free path" $l_{\rm mfp}$ of the conformal fluid is much smaller 
compared to the volume of the fluid which can be taken to be of order one. 
An estimate of $l_{\rm mfp}$ is given by \cite{minwa} 
\be
l_{\rm mfp} \sim \left. \frac{S}{4\pi E} \right|_{\O=0} , 
\ee
where $S$ is the entropy of the black hole. For uncharged black holes, 
$l_{\rm mfp}$ is simply proportional to $1/T$. 
For charged black holes, $l_{\rm mfp}$ depends also on 
the chemical potentials such that 
it is possible to take
an extremal ($T\goto 0$) limit while keeping all 
physical quantities finite. 
In the following sections, 
we will show how the results of \cite{minwa} is 
reflected on the near horizon geometry 
of the extremal black holes.   

\subsection{Extremal black hole in minimal supergravity}

In the minimal gauged supergravity, 
the most general four parameter family of black hole solutions was
obtained in \cite{pope1}. 
We study the extremal limit of the general solution 
and take a close look at the near horizon geometry, 
so that we can use the result of this section as a guide 
when we study more general theories in later sections. 

The solution of \cite{pope1}, with our normalization for $A$ as in 
(\ref{mini}), is given by
\be
ds^2 &=& -\fft{\Delta_\theta\, [(1+g^2 r^2)\rho^2 dt + 2q \nu]
\, dt}{\Xi_a\, \Xi_b \, \rho^2} 
+ \fft{f}{\rho^4}\Big(\fft{\Delta_\theta \, dt}{\Xi_a\Xi_b} -
\omega\Big)^2 + \fft{\rho^2 dr^2}{\Delta_r}
\nn\\
&& +
\fft{\rho^2 d\theta^2}{\Delta_\theta} 
+ \fft{2q\, \nu\omega}{\rho^2} 
+ \fft{r^2+a^2}{\Xi_a}\sin^2\theta d\phi_1^2 + 
      \fft{r^2+b^2}{\Xi_b} \cos^2\theta d\phi_2^2\,,\label{5met}\\
A &=& \fft{q}{\rho^2}\,
         \Big(\fft{\Delta_\theta\, dt}{\Xi_a\, \Xi_b} 
       - \omega\Big)\,,\label{gaugepot}
\ee
where
\be
\rho^2 = r^2 + a^2 \cos^2\theta + b^2 \sin^2\theta\,,
&& \Xi_a = 1-a^2 \,,\quad \Xi_b = 1-b^2 \,,\nn\\
\nu = b\sin^2\theta d\phi_1 + a\cos^2\theta d\phi_2 \,, 
&& \omega = a\sin^2\theta \fft{d\phi_1}{\Xi_a} + 
              b\cos^2\theta \fft{d\phi_2}{\Xi_b}\,,\nn\\
\Delta_\theta = 1 - a^2 \cos^2\theta -
b^2 \sin^2\theta\,,
&& f = 2 m \rho^2 - q^2 + 2 a b q \rho^2\,,
\ee
and
\be
\label{dr2}
\Delta_r = \fft{(r^2+a^2)(r^2+b^2)(1+ r^2) + q^2 +2ab q}{r^2} - 2m 
\,.
\ee
The physical quantities characterizing the solutions are given by \cite{pope1, pope2} :
\be
Q &=& \frac{\pi}{4G} \frac{q}{(1-a^2)(1-b^2)},
\;\;\;\;\;
\mu \,=\,  \frac{3 q r_+^2 }{(r_+^2+a^2)(r_+^2+b^2)+abq },
\\
J_1 &=& \frac{\pi}{4G} \frac{2am+ q b(1+a^2)}{(1-a^2)^2(1-b^2)}, 
\;\;\;\;\; 
\O_1 \,=\, \frac{a(r_+^2+b^2)(r_+^2+1)+bq}{(r_+^2+a^2)(r_+^2+b^2)+abq } ,
\\
J_2 &=& \frac{\pi}{4G} \frac{2bm+ q a(1+b^2)}{(1-b^2)^2(1-a^2)}, 
\;\;\;\;\;
\O_2 \,=\, \frac{b(r_+^2+a^2)(r_+^2+1)+aq}{(r_+^2+a^2)(r_+^2+b^2)+abq } ,
\\ 
S &=& \frac{\pi^2}{2G} \frac{(r_+^2+a^2)(r_+^2+b^2)+abq}{r_+(1-a^2)(1-b^2)} , 
\;\;\;\;\;
T = \frac{r_+^4(1+a^2+b^2+2r_+^2) - (ab + q)^2}{2 \pi r_+ [(r_+^2+a^2)(r_+^2+b^2) +abq]} , 
\label{tbh2}
\\
E &=& \frac{\pi}{4G} \frac{m(3-a^2-b^2-a^2b^2)+2qab(2-a^2-b^2)}{(1-a^2)^2(1-b^2)^2}.
\ee
The zeros of the function $\D_r$ give the locations of the horizons. 
Since $r^2 \D_r$ is a cubic polynomial of $r^2$, we can write 
\be
\label{dr32}
r^2 \D_r = (r^2 -r_+^2) (r^2-r_0^2) (r^2 - r_-^2) . 
\ee
We assume that $r_+^2 \ge r_0^2 \ge r_-^2$ by definition. 
Comparing (\ref{dr2}) and (\ref{dr32}), we note that
\be
r_+^2 + r_0^2 + r_-^2 &=& - (a^2+b^2+1) , 
\nn \\
 r_+^2 r_0^2 + r_0^2 r_-^2 + r_-^2 r_+^2 &=& a^2+b^2 + a^2b^2 - 2m ,
\nn \\
r_+^2 r_0^2 r_-^2 &=& -(ab+q)^2 . 
\label{rpm2}
\ee
We can use (\ref{rpm2}) to show that the numerator of 
the expression for the temperature (\ref{tbh2}) can be rewritten as 
\be
r_+^2 (r_+^2 -r_0^2)(r_+^2 - r_-^2).
\ee
Therefore, in the extremal case, $T=0$, $r_+^2$ must coincide with $r_0^2$, 
as one may have expected on general grounds.

In general, the solution (\ref{5met} - \ref{gaugepot}) 
has four parameters, $(m, q, a,b)$. Extremality imposes a constraint, so
$m$ can be regarded as a function of $a$, $b$ and $q$. 
We find it more convenient to use $(r_+, a,b)$ instead of $(q, a,b)$ 
as independent parameters. 
The reason is that while it is easy to solve (\ref{rpm2}) for $q$ (with $r_0=r_+$),
\be
\label{qrab}
q = r_+^2\sqrt{1+a^2+b^2 +2r_+^2} - ab ,
\ee 
it is difficult to invert this relation to express $r_+$ in terms of $q$ and $a$. 
The parameter $m$ is also easily expressed as a function of $a$, $b$ and $r_+$, 
\be
2m = 3r_+^4+2r_+^2(1+a^2+b^2)+a^2+b^2+a^2 b^2.
\ee

Let us now discuss the allowed values of $(r_+, a, b)$. 
The temperature and entropy are non-negative if and only if $r_+$ is non-negative. 
Without loss of generality, we can assume that $J_a$, $J_b$ and $Q$ 
are all positive. 
It then follows that $a$, $b$ and $q$ should be positive. 
For the angular momenta $J_a$, $J_b$, charge $Q$ and mass $E$ 
to be finite, we also have $a^2, b^2 < 1$. 

It was shown in \cite{pope1} that the black hole is supersymmetric 
if and only if 
\be
\label{rbps}
r_+^2 = a+b+ab .  
\ee
In the space of extremal black holes, there is a sense in which 
BPS black holes are the smallest ones. 
The Euclidean action for the black hole solution was computed in \cite{pope2}:
\be
\label{euca}
I = \frac{\pi}{4G} \frac{T^{-1}}{(1-a^2)(1-b^2)}
\left[
m-(r_+^2+a^2)(r_+^2+b^2) -\frac{q^2 r_+^2}{(r_+^2+a^2)(r_+^2+b^2)+abq} 
\right]. 
\ee
The black hole is thermodynamically stable compared 
to thermal AdS only if the Euclidean action is positive. 
This is analogous to the Hawking-Page phase transition \cite{hp} 
of AdS-Schwarzschild black holes. 
{}From (\ref{euca}), we find that the extremal limit 
of the critical surface $I=0$ coincides with the BPS condition (\ref{rbps}). 
For fixed values of $a$ and $b$, the black hole is stable 
only if $r_+^2$ is greater than or equal to the BPS value (\ref{rbps}). 
\footnote{
Extra care should be taken for BPS black holes. 
It was shown in \cite{silva1, silva2, silva3} that 
the correct thermodynamic variables are not
the BPS values of $\Omega^a$ and $\m$, but their 
next to leading coefficients in a near extremal expansion. 
It was also shown that there is a phase transition in the $(a,b)$ plane. 
We thank P. Silva for drawing our attention to these references.}

\begin{figure}[htb]
\begin{center}
\includegraphics[width=9cm]{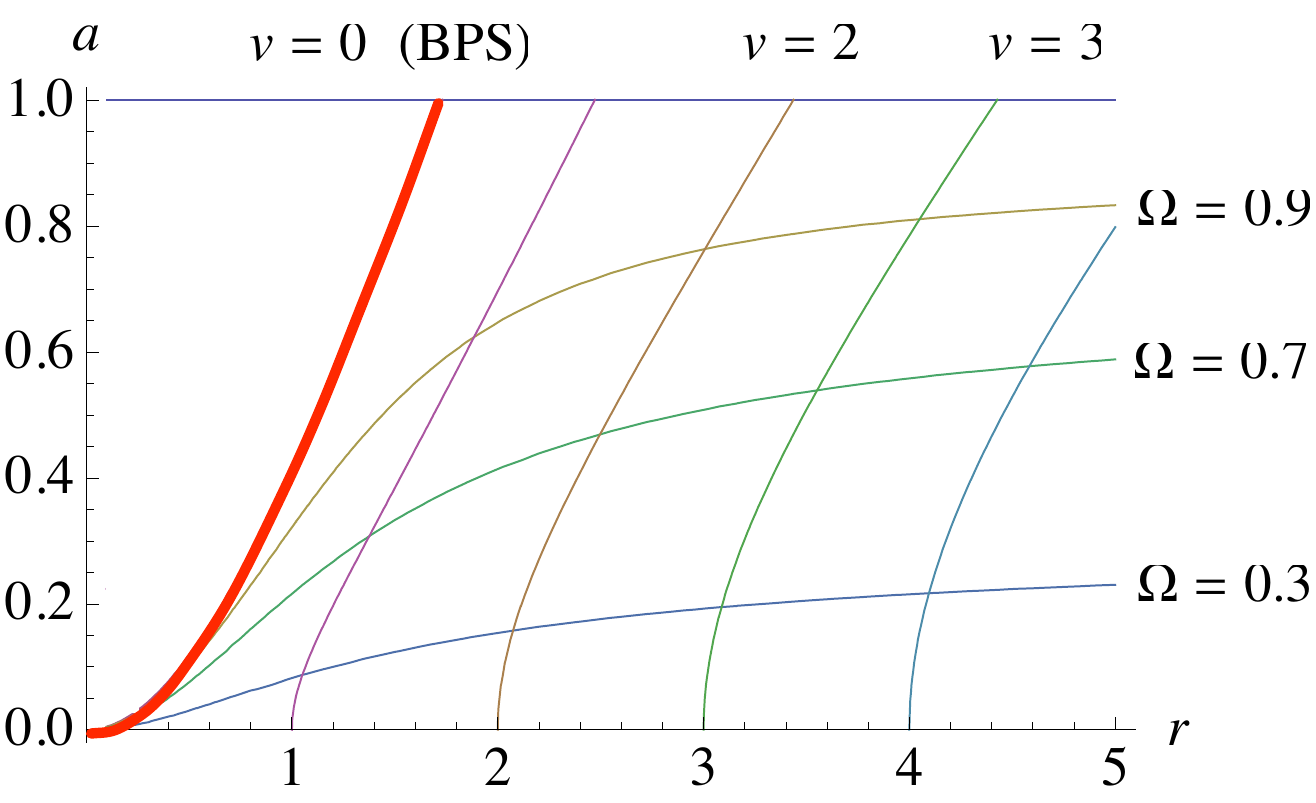}
\caption{Moduli space of extremal black holes.} \label{ra}
\end{center}
\end{figure}

Figure \ref{ra} depicts the $a=b$ slice of the resulting domain. 
Only the region on the right side of the BPS curve is physical.
The value of $q$ in (\ref{qrab}) is positive throughout the physical region. 
The mean free path $l_{\rm mfp}$ is found to be \cite{minwa}, 
for large $r_+$, 
\be
l_{\rm mfp} \sim \left. \frac{S}{4\pi E} \right|_{\O=0} 
\propto \frac{1}{r_+} .  
\ee
Thus we see that, as emphasized in \cite{minwa}, 
the BPS black holes lie on the opposite extreme of 
the ``large black hole limit,'' where 
the fluid mechanics becomes a good approximation. 

One of the prediction from the fluid mechanics is 
that, in the large black hole limit, the entropy takes the form,
\be
S = \frac{\pi}{4G} \frac{2\pi v^3}{(1-\O_1^2)(1-\O_2^2)} . 
\ee
where $v$ is inversely proportional to $l_{\rm mfp}$ 
up to a coefficient of order unity. 
In what follows, we will take $v$ as a measure of 
the effective size of the black hole, 
a convenient notion especially when direct evaluation 
of $l_{\rm mfp}$ is not available. 
Figure \ref{ra} include the contours of fixed values of $v$ 
as well as contours of fixed values of $\O$. 
Note that the effective size of the BPS black holes vanishes, 
although their actual size can be arbitrarily large. 
The reason for the discrepancy is that all BPS black holes have $\O=1$. 

\subsection{Near horizon geometry : a first look}

For simplicity, we begin with the 
case where the two rotation parameters are equal, $a=b$. 
To make the $SU(2)\subset SO(4)$ symmetry manifest, 
we make the usual coordinate change, 
\be
\th = \frac{\tilde{\th}}{2}, 
\;\;\;\;\;
\phi_1 = \frac{\tilde{\psi}-\tilde{\phi}}{2}, 
\;\;\;\;\;
\phi_2 = \frac{\tilde{\psi}+\tilde{\phi}}{2}, 
\ee
and introduce the invariant one-forms, 
\be
\s_1 = \cos\tilde{\psi} d\tilde{\th} + \sin\tilde{\psi} \sin\tilde{\th} d\tilde{\phi}, 
\;
\s_2 = - \sin\tilde{\psi} d\tilde{\th} + \cos\tilde{\psi} \sin\tilde{\th} d\tilde{\phi}, 
\;
\s_3 = d\tilde{\psi} + \cos\tilde{\th} d\tilde{\phi} .
\ee 
The near horizon limit of the extremal solution 
takes a remarkably simple form:
\be
ds^2 &=& v_1^2 \left( -u^2 dt^2 + \frac{du^2}{u^2} \right) + 
v_2^2 \left[ (\s_1)^2 + (\s_2)^2 \right] + v_3^2 (\s_3 +\a\, u dt)^2 ,
\\
A &=& e u dt + b \s_3 ,
\ee
where the near horizon parameters are given by
\be
v_1^2 = \frac{r_+^2+a^2}{4(3r_+^2+2a^2+1)}, \;\;\;
v_2^2 = \frac{r_+^2+a^2}{4(1-a^2)} ,  \;\;\;
v_3^2 = \left( \frac{(r_+^2+a^2)^2+a^2q}{2 r_+ (r_+^2+a^2)(1-a^2)} \right)^2 , 
\\
e = \frac{r_+ q}{2(r_+^2+a^2)(3r_+^2+2a^2+1)}  ,  \;\;\;
b = \frac{aq}{2(r_+^2+a^2)(1-a^2)}, 
\\
\a = \frac{a(1-a^2)}{r_+(3r_+^2+2a^2+1)}
\left[1+\frac{2r_+^2 q}{(r_+^2+a^2)^2+a^2 q} \right].  
\ee
The large black hole limit in the sense discussed above 
is obtained by taking 
$r_+\goto \infty$ while keeping $a$ fixed. 
In this limit, $q \goto \sqrt{2} r_+^3$, and the near horizon data reduce to
\be
v_1^2 =\frac{1}{12}, 
&
\displaystyle{v_2^2=\frac{r_+^2}{4(1-a^2)}}, 
& v_3^2 = \frac{r_+^2}{4(1-a^2)^2},
\\
e=\frac{\sqrt{2}}{6},
& 
\displaystyle{b=\frac{\sqrt{2}a r_+}{2(1-a^2)}}, 
&
\a = \frac{2\sqrt{2}a(1-a^2)}{3r_+^2} . 
\ee
The physical quantities also simplify:
\be
Q = \frac{\pi}{4G} \frac{\sqrt{2}r_+^3}{(1-a^2)^2},
\;\;\;
& \displaystyle{J = \frac{\pi}{4G} \frac{3ar_+^4}{(1-a^2)^3},}&
\;\;\;
S = \frac{\pi}{4G} \frac{2\pi r_+^3}{(1-a^2)^2} ,
\\
\m =  3\sqrt{2} r_+,
\;\;\;
&\displaystyle{\O = a,}&
\;\;\;
E = \frac{\pi}{4G} \frac{3(a^2+3)r_+^4}{2(1-a^2)^3}.
\ee

In the general case with $a\neq b$, 
the near horizon limit for arbitrary values of $(r_+,a,b)$ 
is quite messy, so we focus on the large black hole limit. 
To the leading order in $r_+$, the near horizon solution is given by 
($c_\th \equiv \cos\th$, $s_\th \equiv \sin\th$)
\be
\label{sol2met}
ds^2 &=& \frac{1}{12}\left(-u^2dt^2 + \frac{du^2}{u^2} \right) 
+ \frac{r_+^2 d\th^2}{1-a^2 c_\th^2-b^2s_\th^2} 
+ r_+^2 G_{ab}(\th) D\phi^a D\phi^b,
\\
\label{sol2gau}
A &=& \tilde{e} \,udt + b_a(\th) D\phi^a
\;\;\;\;\; \;\;\;
\left( D\phi^a \equiv d\phi^a+\a^a u dt \right).
\ee
The $U(1)^2$ fibration part of the metric is rather non-tivial:
\be
G_{ab} = (L^T \hat{G} L)_{ab} ,
\;\;
\hat{G} = 
\begin{pmatrix}
1-a^2c_\th^2 & abs_\th c_\th \\ abs_\th c_\th & 1-b^2s_\th^2 
\end{pmatrix},
\;\;
L = \text{diag}\left(\frac{s_\th}{1-a^2}, \frac{c_\th}{1-b^2}\right).
\label{sol2met2}
\ee
The other parameters in the solution are given by
\be
(\a^1,\a^2) &=& \frac{\sqrt{2}}{3r_+^2}(b(1-a^2),a(1-b^2)),
\\
(b_1, b_2) &=& \sqrt{2}r_+ 
\left( \frac{a s_\th^2}{1-a^2}, \frac{b c_\th^2}{1-b^2} \right) ,
\;\;\; \tilde{e} = \frac{\sqrt{2}}{6} .
\ee
The physical quantities in the large black hole limit are given by
\be
Q = \frac{\pi}{4G} \frac{\sqrt{2}r_+^3}{(1-a^2)(1-b^2)}, 
&& \mu =  3\sqrt{2} r_+, 
\\
J_1 = \frac{\pi}{4G} \frac{3ar_+^4}{(1-a^2)^2(1-b^2)},
&& \O_1 = a,
\\
S = \frac{\pi}{4G} \frac{2\pi r_+^3}{(1-a^2)(1-b^2)} , 
&&
E = \frac{\pi}{4G} \frac{3(3-a^2-b^2-a^2b^2)r_+^4}{2(1-a^2)^2(1-b^2)^2} . 
\ee

\section{Extremal black holes - I. equal rotation}

We now begin our analysis of the near horizon geometry 
of black hole solutions in the class of supergravity 
theories reviewed in section 2. 
In this section, we will focus on the equal rotation case, 
in which the $SU(2)$ isometry simplifies the computation drastically, 
relegating the unequal rotation case to the next section. 

\subsection{Entropy function}

The general near horizon solution of an extremal black hole in AdS$_5$, 
which has an $SO(2,1)\times SU(2)$ invariance, take the following form:
\footnote{
In the special case of the $U(1)^3$ theory, 
the entropy function for rotating AdS$_5$ black holes 
was previously obtained in \cite{sam, silva4}. 
Here, we consider the large class of gauged supergravity theories all at once.
}
\be
\label{nhag}
ds^2 &=& v_1^2 \left( -u^2 dt^2 + \frac{du^2}{u^2} \right) + 
v_2^2 \left[ (\s_1)^2 + (\s_2)^2 \right] + v_3^2 (\s_3 +\a\, u dt)^2 ,
\\
\label{nhav}
A^I &=& e^I u dt + b^I \s_3 \; = \;  \tilde{e}^I u dt + b^I (\s_3 +\a u dt), \;\;  
F^I = e^I du\wedge dt - b^I \s_1 \wedge \s_2,
\\
\label{nhas}
X^I &=& u^I
\ee
We follow the standard procedure of the entropy function formalism 
\cite{sen1,sen2,sen5}.
First, we integrate the Lagrangian over the horizon to 
obtain the ``near horizon action'' : 
\be
\CF &\equiv& \frac{\pi}{G} v_1^2 v_2^2 v_3 
\left(-\frac{2}{v_1^2} +\frac{2}{v_2^2} -\frac{v_3^2}{2v_2^4}
+\frac{v_3^2\a^2}{2v_1^4} 
- 2V +\frac{Q_{IJ}e^Ie^J}{2v_1^4} - \frac{Q_{IJ}b^Ib^J}{2v_2^4} \right)
\nn \\
&& - \frac{\pi}{6G} C_{IJK} \left(3\tilde{e}^I + 2\a b^I \right) b^J b^K .
\ee
The Chern-Simons term is slightly subtle. 
Inserting the ansatz (\ref{nhav}) with constant 
$\tilde{e}^I$, $b^I$ into the Lagrangian yields an incorrect result. 
To obtain the correct answer, one should consider $b^I$ as a function of $u$, 
integrate by parts, and set $b^I$ to be constant at the final stage.\footnote{
We thank S. Trivedi and K. Goldstein for clarifying this point.}
See appendix A for details.

The near horizon equations of motion can be derived by extremizing,
$\CF$ 
\be
\frac{\partial \CF}{\partial v_i}=0, 
\;\;\;\;\;  
\frac{\partial \CF}{\partial b^I} =0, 
\;\;\;\;\;
\frac{\partial \CF}{\partial u^I} =0,
\ee
while keeping $(\tilde{e}^I,\a)$ fixed. Explicitly, the equations read 
as follows.
\bn

\item
Einstein equation:
\be
\label{ein1a}
\frac{1}{v_1^2}+\frac{v_3^2}{2v_2^4}-\frac{v_3^2\a^2}{v_1^4} 
&=& \frac{Q_{IJ}e^Ie^J}{2v_1^4} ,
\\
\label{ein1b}
\frac{1}{v_2^2}-\frac{v_3^2}{v_2^4} + \frac{v_3^2\a^2}{2v_1^4} 
&=& \frac{Q_{IJ}b^Ib^J}{2v_2^4},
\\
\label{ein1c}
-\frac{1}{v_1^2}+\frac{1}{v_2^2} &=& 2V.
\ee

\item
Maxwell equation:
\be
\label{max1}
Q_{IJ}\left[ (v_1^{-2}v_2^2v_3)(e^J \a -v_1^4v_2^{-4}b^J) \right] 
- C_{IJK}e^Jb^K = 0. 
\ee

\item
Scalar equation:
\be
\label{sc1}
\frac{\partial Q_{IJ}}{\partial \vph^i} 
\left(v_1^{-4}e^I e^J - v_2^{-4}b^I b^J \right) 
- 4 \frac{\partial V}{\partial \vph^i} =0.
\ee

\en

\noindent
Note that the scalar equation incorporates 
the constraint, $\frac{1}{6}C_{IJK}u^Iu^Ju^K=1$, 
by means of a Lagrange multiplier.
It is straightforward to confirm that the same equations follow from   
writing down the full equations of motion and inserting the near horizon ansatz. 

The electric charges and the angular momentum are 
the conjugate variables of $(\tilde{e}^I, \a)$ with respect to $\CF$, 
\be
\label{qn}
Q_I &=& \frac{\partial \CF}{\partial \tilde{e}^I}
\; = \; \frac{\pi}{G}\left[
(v_1^{-2}v_2^2 v_3)  Q_{IJ} e^J -\half C_{IJK}b^Jb^K \right] ,
\\
J &=& \frac{\partial \CF}{\partial \a} 
\; = \; \frac{\pi}{G} \left[ (v_1^{-2}v_2^2v_3) 
( v_3^2 \a + Q_{IJ}b^I e^J) - \frac{1}{3} C_{IJK} b^I b^J b^K \right].
\label{jn}
\ee
An alternative method to evaluate $Q_I$ and $J$ by using only the 
near horizon data in five dimensional supergravity was given in \cite{sur3,tach2}
($i_\xi \s_3=-1$),
\be
\label{qq}
Q_I &=& \frac{1}{16\pi G} \int_{S^3_{\rm hor}} 
\left( Q_{IJ} \star F^J + \thalf C_{IJK} A^J \wedge F^K \right) , 
\\
J &=& -\frac{1}{16\pi G} \int_{S^3_{\rm hor}}  \left[ \star d\xi 
+ (i_\xi A^I) \left(Q_{IJ} \star F^J 
+ \textstyle{\frac{1}{3}} C_{IJK} A^J \wedge F^K \right) \right] . 
\label{jj}
\ee
It is easy to check that the two methods give the same results.

The entropy function is the Legendre transform of the near horizon action:
\be
\label{ent1}
\CE &\equiv& Q_I \tilde{e}^I +J\a - \CF 
\nn \\
&=& \frac{\pi}{G}
v_1^2 v_2^2 v_3 \left(\frac{2}{v_1^2} - \frac{2}{v_2^2} + \frac{v_3^2}{2v_2^4}
+\frac{v_3^2\a^2}{2v_1^4} +2V  
+ \frac{Q_{IJ}e^Ie^J}{2v_1^4} + \frac{Q_{IJ}b^Ib^J}{2v_2^4} \right).
\ee
To complete the Legendre transformation, 
the variables $(e^I,\a)$ in $\CE$ should be eliminated 
in favor of the conjugates $(Q_I,J)$. Before doing so, 
we note that applying the Einstein equations (\ref{ein1a}-\ref{ein1c}) 
to (\ref{ent1}) yields the area law as expected;
\be
2\pi \CE = \frac{4\pi^2}{G} v_2^2 v_3 = \frac{\text{(Area)}_{\text{hor}}}{4G}
=S_{\text{BH}}.
\ee

\subsection{Large black hole solutions}

Motivated by the large black hole limit 
of the minimal supergravity discussed in the previous section, we set
\be
v_2^2 \equiv \frac{v^2}{4(1-a^2)}, \;\;\;\;\; 
v_3^2 \equiv \frac{v^2}{4(1-a^2)^2}, 
\ee
and consider the limit $v \gg 1$ with $a$ fixed.
We then try to solve the near horizon equations 
by expanding the other variables in powers of $1/v$. 

\subsubsection{Minimal supergravity revisited}

As a warm up exercise, let us see how the minimal supergravity 
solution can be recovered from the near horizon equations. 
Setting $e^I = e \bar{X}^I$, $b^I = b \bar{X}^I$ 
in (\ref{ein1a}-\ref{max1}), the equations reduce to 
\be
\label{einma}
\frac{1}{v_1^2}+\frac{v_3^2}{2v_2^4}-\frac{v_3^2\a^2}{v_1^4} 
&=& \frac{3e^2}{2v_1^4} ,
\\
\label{einmb}
\frac{1}{v_2^2}-\frac{v_3^2}{v_2^4} + \frac{v_3^2\a^2}{2v_1^4} 
&=& \frac{3b^2}{2v_2^4},
\\
\label{einmc}
\frac{1}{v_1^2} - \frac{1}{v_2^2} &=& 12.
\\
\label{maxm}
(v_1^{-2}v_2^2v_3)(e \a -v_1^4v_2^{-4}b)  &=& 2 e b . 
\ee
We can eliminate the $\a$-dependent terms in (\ref{einma}) and (\ref{einmb}),
\be
\frac{1}{v_1^2} +\frac{2}{v_2^2}-\frac{3v_3^2}{2v_2^4} = 
\frac{3e^2}{2v_1^4}+\frac{3b^2}{v_2^4}.
\ee 
Since the left hand side is $\CO(1)$ in the leading order, we have two possibilities:
\be
\text{(a) } e \sim \CO(1), \;\; b \lesssim \CO(v^2) , \;\;\;\;\; 
\text{(b) } e \lesssim \CO(v^{-1}), \;\; b \sim \CO(v^2).
\ee
For the possibility (b), we find from (\ref{einma}) and (\ref{einmb}) that
\be
\label{unph}
\a = \pm \frac{v_1}{v_3}, \;\;\;\;\; 
b = \pm \frac{v_2^2}{\sqrt{3}v_1} .
\ee
This ``solution'' cannot be the near horizon geometry of a rotating black hole.  
As $a\goto 0$, the black hole horizon becomes a round three-sphere, 
and we expect the black hole to become static. 
But, in (\ref{unph}) the rotation parameter $\a$ 
and the magnetic dipole moment $b$ retain non-zero values 
even when $a$ vanishes.

The true solution, which comes from the possibility (a), can be summarized as follows:
\be
v_1^2 =\frac{1}{12}, 
&
\displaystyle{v_2^2=\frac{v^2}{4(1-a^2)}}, 
& v_3^2 = \frac{v^2}{4(1-a^2)^2},
\\
e=\frac{\sqrt{2}}{6},
& 
\;\;\; \displaystyle{b=\frac{\sqrt{2}a v}{2(1-a^2)}}, \;\;\;
&
\a = \frac{2\sqrt{2}a(1-a^2)}{3 v^2} .
\ee
The physical quantities also can be easily computed:
\be
Q = \frac{\pi}{4G} \frac{\sqrt{2}v^3}{(1-a^2)^2},
\;\;\;
& \displaystyle{J = \frac{\pi}{4G} \frac{3av^4}{(1-a^2)^3},}&
\;\;\;
S = \frac{\pi}{4G} \frac{2\pi v^3}{(1-a^2)^2} .
\ee
Of course, all the results agree perfectly with the large, extremal limit 
of the general solution we reviewed earlier. 

\subsubsection{General solution and universality}

Since all supergravity theories contain the minimal supergravity, 
we expect the same $v$ dependence of the near-horizon parameters 
in the large black hole limit. 
To the leading order in $1/v$, the equations take the following form.
\be
\label{einga}
2v_1^2 \;= \; - \frac{1}{V} &=& Q_{IJ}e^Ie^J,
\\
\frac{\partial Q_{IJ}}{\partial \vph^i} e^I e^J 
&=& \frac{1}{V^2}\frac{\partial V}{\partial \vph^i} ,
\label{scg}
\\
\label{eingb}
\frac{v_2^2 v_3^2 \a^2}{2v_1^4} - \frac{Q_{IJ} b^I b^J}{2v_2^2}
&=&
\frac{v_3^2}{v_2^2} - 1 ,
\\
\label{maxg}
\frac{v_2^2v_3}{v_1^2} Q_{IJ} e^J \a
&=& C_{IJK}e^J b^K .
\ee
Guided by the minimal supergravity solution, we introduce the reparametrization,   
\be
\label{repar}
\a = \left(\frac{\bar{\a}v_1^2}{v^2}\right) 8\sqrt{2} a(1-a^2) , \;\;\;\;\;  
b^I = \bar{\a} \bar{b}^I \frac{\sqrt{2}a v}{2(1-a^2)} .
\ee
In terms of the new variables, the equations (\ref{eingb}, \ref{maxg}) read,
\be
\label{new1}
\bar{\a}^2(4 - Q_{IJ} \bar{b}^I \bar{b}^J) &=& 1, 
\\
\label{new2}
Q_{IJ} e^J  - \thalf C_{IJK} e^J \bar{b}^K &=&0,
\ee
which are completely independent of the parameter $a$. 
The other two equations (\ref{einga}, \ref{scg}) are also independent of $a$. 
We conclude that 
the near horizon geometry of the large black holes in AdS$_5$ 
has a universal dependence on $v$ and $a$, 
in agreement with the prediction of the fluid mechanics 
of the dual CFT \cite{minwa}. 

Let us now discuss how to solve the remaining equations. 
In a theory with $n$ vector fields, 
the extremal black hole solution (with equal rotation) 
depends on $n+1$ parameters. 
Physically, they correspond to $n$ charges and the angular momentum. 
In solving the near-horizon equations, 
we will choose the parameters to be 
$v$, $a$ and the $n-1$ independent values of $\vph^i$.

In principle, it is easy to see how to solve the above equations. 
First, (\ref{einga}) and (\ref{scg}) give $n$ quadratic equations for $\{e^I\}$, 
so that we can solve them to express $\{e^I\}$ as functions of $\{\vph^i\}$. 
Second, (\ref{new2}) gives $n$ linear equations for $\bar{b}^I$ in terms of 
$\{e^I\}$ (and $\vph^i$ through $Q_{IJ}$). 
Finally, (\ref{new1}) determines $\bar{\a}$ in terms of $\bar{b}^I$. 

In practice, the complete solution can be quite complicated 
because, in general, the equations for $e^I$ lead to 
a polynomial equation of degree $2n$. 
Some theories may have extra symmetries to simplify the problem. 
Otherwise, one could look for a closed sub-family of solutions 
which effectively lowers the value of $n$. We will discuss such an example 
shortly.

\subsubsection{$U(1)^3$ theory}

To illustrate the algorithm to solve the near-horizon equations, 
we consider the $U(1)^3$ theory with $C_{123}=1$. 
This theory has many simplifying features such as 
$u_I = (u^I)^{-1}$, 
$Q_{IJ} = (u_I)^2 \d_{IJ}$ and $V = -2(u_1+u_2+u_3)$.  
The scalar equations can be written as 
\be
2u_I + 2V^2 u_I^2 (e^I)^2 = -V 
\ee
for each $I$ (no sum). It is straightforward to solve all the equations, 
and the answer can be written as
\be
v_1^2 = \frac{\left( L_1L_2L_3 \right)^{-1/3}}{4\left(\sum_I L_I^{-1}\right)},
&&
\bar{\a} = \sqrt{\prod_I \frac{(1-L_I)}{2 L_I}}  , 
\\
e^I = \frac{\sqrt{L_I^3(1-L_I)}}{2\left(\sum_I L_I^{-1}\right)L_1L_2L_3}, 
&&
\bar{b}^I = \frac{\sqrt{L_I(1-L_I)}}{\bar{\a}\sqrt{2}(L_1L_2L_3)^{1/3}}. 
\ee
Here, the variables $L_I$ satisfy $L_I > 0$ and $ L_1+L_2+L_3=1$ 
(they are the same as $X_i$ defined in section 6.2. of \cite{minwa}). 
This parametrization makes it clear that the physical quantities
\be
Q_I = \frac{\pi}{4G} \sqrt{\frac{1-L_I}{L_I}} \frac{v^3}{(1-a^2)^2},
\;\;\;\;\;
J = \frac{\pi}{4G} \frac{1}{(L_1L_2L_3)^{1/3}} \frac{2av^4}{(1-a^2)^3},
\ee
match perfectly with the $T\goto 0$ limit of 
the results found in \cite{pope--,pope-,minwa}.
\footnote{
To compare, for instance, our result with eq. (75) of \cite{minwa}, 
note that $\pi/4G = N^2/2$ in this theory and 
that $v$ here is identified with $2\pi T(XYZ)^{1/3}/(X+Y+Z-1)$ there.}

\subsubsection{A new solution} 

Next we consider an example of a theory 
of which no black hole solution is known in the literature.  
It has the following non-vanishing components of $C_{IJK}$, 
\be
C_{123}=C_{234}=C_{341}=C_{412} = \frac{1}{4} .
\ee 
This is the abelian truncation of the supergravity 
dual to the famous conifold CFT of \cite{kw}. 
The four $U(1)$ symmetries correspond to $U(1)_R$ symmetry, 
two mesonic symmetries and one baryonic symmetry. 
The scalars can be decomposed according to 
which vector multiplet they belong to:
\be
X^1 = r+t+s_1, \;\; X^2= r-t+s_2, \;\; 
X^3 = r+t-s_1, \;\; X^4= r-t-s_2. \;\; 
\ee
The discrete symmetries of the theory 
makes it consistent to turn off the mesonic charges and 
relevant scalars identically, $s_1=s_2=0$.  
The constraint $\frac{1}{6}C_{IJK}X^IX^JX^K=1$ then becomes 
\be
r^3-r t^2 = 1\;\;\; \imp \;\;\ t = \sqrt{r^2-\frac{1}{r}} \;\; ,
\ee
where $t$ is taken to be positive without loss of generality.

Following the general procedure described in the previous subsection, 
we first solve 
\be
\label{coni-e}
Q_{IJ} e^I e^J = -\frac{1}{V}, \;\;\;\;\;
(\p_r Q_{IJ}) e^I e^J = -\p_r\left(\frac{1}{V}\right) .
\ee
In the $(r,t)$ basis, $e^I=(e_r,e_t)$, 
\be
Q_{IJ} &=& 
\begin{pmatrix}
Q_{rr} & Q_{rt} \\
Q_{tr} & Q_{tt} 
\end{pmatrix}
=
\begin{pmatrix}
4r^4-2r+r^{-2} & -4r^3 t \\
 -4r^3 t & 4r^4-2r
\end{pmatrix}
\\
\p_r Q_{IJ} &=& 
\begin{pmatrix}
16r^3-2-2r^{-3} & -2rt^{-1}(8r^3-5) \\
-2rt^{-1}(8r^3-5) & 16r^3-2
\end{pmatrix}
\\
-\frac{1}{V}
&=& \frac{4r^3-1}{18r^5} ,
\;\;\;\;\;
-\p_r\left(\frac{1}{V}\right)
= - \frac{8r^3-5}{18r^6} .
\ee
Equations (\ref{coni-e}) can be solved for $(e_r, e_t)$ in a closed form, 
though the answer is rather complicated. 
The next step is to solve (\ref{new2}) : 
\be
Q_{IJ} e^J  - \thalf C_{IJK} e^J \bar{b}^K =0
\;\;\; \imp \;\;\; 
\begin{pmatrix}
3 e_r& - e_t \\
-e_t & -e_r 
\end{pmatrix}
\begin{pmatrix}
\bar{b}_r \\ \bar{b}_t 
\end{pmatrix}
= 
\begin{pmatrix}
Q_{rr} & Q_{rt} \\
Q_{tr} & Q_{tt} 
\end{pmatrix}
\begin{pmatrix}
e_r \\ e_t 
\end{pmatrix}.
\ee
Again, the answer for $(\bar{b}_r, \bar{b}_t)$ can be written down 
explicitly. Finally, (\ref{new1}) yields $\bar{\a}$ and 
complete the solution. 

Although the explicit expressions for the 
near horizon solution are not very illuminating, 
this example clearly shows how the general method describe above 
can be used to solve all the near horizon equations.


\section{Extremal black holes - II. Unequal rotation}

\subsection{Entropy function}

Following \cite{sen2, kevin}, we write down the near horizon ansatz 
for the most general rotating black holes in AdS$_5$.
\footnote{The entropy function of general rotating black holes
in 5$d$ ungauged supergravity was first considered in \cite{kevin}. 
To our knowledge, the results in gauged supergravity presented here are new, 
apart from the general near horizon analysis for {\em supersymmetric} 
black holes performed from a different angle in \cite{kun1,kun2}.}
\be
\label{j2met}
ds^2 &=& w_1^2(\th) \left( - u^2 dt^2 + \frac{du^2}{u^2}\right) 
+w_2^2(\th) d\th^2 
+  G_{ab}(\th) 
(d\phi^a + \a^a u dt) (d\phi^b + \a^b u dt),
\\ 
\label{j2gau}
A^I &=& \tilde{e}^I u dt + b^I_a(\th) (d\phi^a + \a^a u dt),
\\
\label{j2sca}
X^I &=& u^I(\th).
\ee
In the metric, we have two functions $w_1, w_2$ and three components of $G_{ab}$. 
One combination out of these five can be removed by reparametrization of $\th$. 
The ansatz preserves $SO(2,1)$ symmetry realized by the Killing vectors,
\footnote{
It was proven in \cite{kun3} that the near horizon geometry of 
any extremal black hole in 
four and five dimensions, in a generic second order theory 
of gravity coupled 
to uncharged scalars and gauge fields (including a cosmological constant) 
must have $SO(2,1)$ symmetry.}
\be
L_{+1} = \partial_t, \;\;\; 
L_0 = t\partial_t - u \partial_u, \;\;\;
L_{-1} = \thalf(1/u^2+t^2)\partial_t-(tu)\partial_u - (\a^a/u) \partial_{\phi^a}.
\ee

In computing the near horizon action, 
the Einstein-Hilbert term deserves some comments: 
\be
R &=& R_1 + R_2 + R_3,
\\
R_1 &=& -\frac{2}{w_1^2} - \frac{2\dot{w}_1^2}{w_1^2w_2^2} 
+\frac{4\dot{w}_1 \dot{w}_2}{w_1 w_2^3}
- \frac{4\ddot{w}_1}{w_1 w_2^2} , 
\;\;\;\;\; 
R_2 \; = \; \frac{G_{ab}\a^a \a^b}{2 w_1^4} ,
\\
R_3 &=& 
\frac{(G^{ab}\partial_\th G_{ba})^2-(G^{ab}\partial_\th G_{bc})(G^{cd} \partial_\th G_{da})}{4w_2^2} 
-\frac{2\partial_\th (w_1^2 w_2^{-1} \partial_\th \sqrt{G})}
{\sqrt{G} w_1^2 w_2} .
\ee
The metric ansatz (\ref{j2met}) takes the form of a two-torus ($\phi^1,\phi^2$) 
fibered over the ``base" space ($t,u,\th$).
The first term is simply the scalar curvature of the base space, 
where dots denote derivatives in $\th$.
The last term is the contribution from the fiber metric. 
The second term can be thought of as a ``Kaluza-Klein" electric flux density. 
Combining with
\be
\sqrt{-g} &=& w_1^2 w_2 \sqrt{G} ,
\ee
we find
\be
\sqrt{-g} R &=& w_1^2 w_2 \sqrt{G} \left(
-\frac{2}{w_1^2} +\frac{2\dot{w}_1^2}{w_1^2w_2^2} 
+\frac{4\dot{w}_1 \partial_\th \sqrt{G}}{w_1 w_2^2 \sqrt{G}}
+ \half G_{ab} \a^a\a^b \right)
\nn \\
&&+ w_1^2w_2\sqrt{G} 
\left( 
\frac{1}{w_2^2}\left(\frac{\partial_\th \sqrt{G}}{\sqrt{G}}\right)^2
+\frac{\partial_\th G^{ab} \partial_\th G_{ab}}{4w_2^2}   \right)
- 2 \partial_\th \left( \frac{1}{w_2} \partial_\th \left(w_1^2\sqrt{G} \right) \right) .
\label{td1}
\ee
The last term, being a total derivative, does not affect the equations of motion. 
However, we will show that it makes a non-zero contribution to the entropy function.

Adding the matter contributions 
and evaluating the Chern-Simons term, we obtain the near horizon action,
\footnote{
In this subsection, 
we denote the Newton's constant by $G_5$ 
to avoid confusion with $G\equiv \det(G_{ab})$.
}
\be
\CF &=& \frac{\pi}{4G_5}\int d\th \sqrt{G}  w_1^2 w_2 
\left( -\frac{2}{w_1^2} -2V 
+ \frac{ G_{ab} \a^a\a^b}{2w_1^4} \right) 
-\frac{\pi}{4G_5} \left[ \frac{2}{w_2}\partial_\th \left(w_1^2\sqrt{G}\right)\right]_0^{\pi/2} 
\nn \\
&& + \frac{\pi}{4G_5}\int d\th \sqrt{G}  w_1^2 w_2 
\left( \frac{2\dot{w}_1^2}{w_1^2w_2^2} 
+\frac{4\dot{w}_1 \partial_\th \sqrt{G}}{w_1 w_2^2 \sqrt{G}}
+ \frac{1}{w_2^2}
\left(\frac{\partial_\th \sqrt{G}}{\sqrt{G}}\right)^2
+\frac{\partial_\th G^{ab} \partial_\th G_{ab}}{4w_2^2} 
\right)
\nn \\
&&+ \frac{\pi}{4G_5}\int d\th \sqrt{G}  w_1^2 w_2  \left( 
\frac{Q_{IJ} e^I e^J }{2w_1^4} 
- \frac{Q_{IJ} G^{ab}  \partial_\th b_a^I \partial_\th b_b^J }{2w_2^2} 
-\frac{g_{ij} \partial_\th \vph^i \partial_\th \vph^j}{2w_2^2} 
\right)
\nn \\
&&- \frac{\pi}{4G_5}\int d\th \left( 
\frac{1}{6} C_{IJK} (3 \tilde{e}^I + 2\a^a b_a^I)\e^{bc}  b_b^J \partial_\th b_c^K 
\right), 
\ee
where $\e^{12}=-1=-\e^{21}$. 
Upon taking a variation, we obtain the near horizon equations:
\bn

\item 
Einstein equation

\bn

\item
$\displaystyle{\frac{w_1}{4} \frac{\d \CF}{\d w_1}+\frac{w_2}{2} \frac{\d \CF}{\d w_2}=0} :$

\be
\label{einx1}
\frac{1}{w_1^2} + 2V  + 
\frac{1}{\sqrt{G}w_1^2w_2} \partial_\th \left( \frac{w_1}{w_2} \partial_\th \left(\sqrt{G}w_1\right) \right) 
&=& 0. 
\label{td2}
\ee

\item
$\displaystyle{\frac{w_1}{4} \frac{\d \CF}{\d w_1}+\frac{w_2}{2} \frac{\d \CF}{\d w_2}
+ G^{ab} \frac{\d \CF}{\d G^{ab}}=0} :$

\be
\label{einx2}
-\frac{1}{w_1^2}
-\frac{\dot{w}_1^2}{w_1^2w_2^2} 
- \frac{3}{w_1w_2}\partial_\th \left(\frac{\dot{w}_1}{w_2}\right) 
+\frac{\dot{w}_1\partial_\th \sqrt{G} }{w_2^2w_1\sqrt{G}}
+\frac{1}{w_2^2}\left(\frac{\partial_\th\sqrt{G}}{\sqrt{G}}\right)^2
 &&
\nn \\
+\frac{\partial_\th G^{ab} \partial_\th G_{ab}}{4w_2^2}
+ \frac{ G_{ab} \a^a\a^b}{w_1^4}
+\frac{Q_{IJ} e^I e^J }{2w_1^4}  
-\frac{g_{ij} \partial_\th \vph^i \partial_\th \vph^j}{2w_2^2} 
= 0. &&
\ee

\item
$\displaystyle{w_2 \frac{\d \CF}{\d w_2}+ G^{ab} \frac{\d \CF}{\d G^{ab}}=0} :$

\be 
-\frac{4}{w_1w_2} \partial_\th\left(\frac{\dot{w}_1}{w_2} \right)
+\frac{2\dot{w}_1\partial_\th\sqrt{G}}{w_2^2 w_1 \sqrt{G}}
-\frac{1}{w_2} \partial_\th \left( \frac{\partial_\th\sqrt{G}}{w_2\sqrt{G}} \right)
+\frac{1}{w_2^2}\left(\frac{\partial_\th\sqrt{G}}{\sqrt{G}}\right)^2
&&
\nn \\
+\frac{\partial_\th G^{ab} \partial_\th G_{ab}}{2w_2^2}
+\frac{ G_{ab} \a^a\a^b}{2w_1^4} - 
\frac{Q_{IJ} G^{ab}  \partial_\th b_a^I \partial_\th b_b^J }{2w_2^2} 
-\frac{g_{ij} \partial_\th \vph^i \partial_\th \vph^j}{w_2^2} 
 = 0.&&
\label{einx3}
\ee

\item 
$\displaystyle{\frac{\d \CF}{\d G^{ab}} - \half \left(G^{cd} \frac{\d \CF}{\d G^{cd}}\right) G_{ab}=0} :$

\be
X_{ab} & -& \half (G^{cd}X_{cd})G_{ab} \; = \; 0,
\nn \\
X_{ab} &\equiv& \frac{1}{\sqrt{G}w_1^2w_2} 
\left[   
\partial_\th\left( \frac{\sqrt{G}w_1^2}{w_2} \partial_\th G_{ab} \right) 
- G_{ac} \partial_\th\left( \frac{\sqrt{G}w_1^2}{w_2}  \partial_\th G^{cd} \right) G_{db} 
\right]
\nn \\
&&+ \frac{2(G_{ac}\a^c)(G_{bd}\a^d)}{w_1^4} 
+ \frac{2 Q_{IJ} \partial_\th b_a^I \partial_\th b_b^J}{w_2^2} .
\label{einx4}
\ee

\en

\item 
Maxwell equation
\be
\frac{\sqrt{G} w_2}{w_1^2} Q_{IJ} e^J \a^a 
+ \partial_\th \left( \frac{\sqrt{G}w_1^2}{w_2} Q_{IJ} G^{ab} \partial_\th b_b^J  \right) 
-C_{IJK} e^J \e^{ab} \partial_\th  b_b^K  = 0.
\ee

\item 
Scalar equation
\be
\frac{2}{ \sqrt{G}w_1^2w_2} \partial_\th 
\left(\frac{\sqrt{G} w_1^2  }{w_2} g_{ij} \partial_\th \vph^j \right) 
+ \frac{\partial Q_{IJ}}{\partial \vph^i} 
\left(\frac{e^Ie^J}{w_1^4} - 
\frac{G^{ab} \partial_\th b^I_a \partial_\th b^J_b}{w_2^2} \right) 
-4 \frac{\partial V}{\partial \vph^i}
= 0. 
\ee

\en
The conserved charges are given by 
\be
\label{qn2}
Q_I &=& \frac{\partial \CF}{\partial \tilde{e}^I}
 = 
\frac{\pi}{4G_5} \int d\th  \left( 
\frac{\sqrt{G}w_2}{w_1^2} Q_{IJ} e^J - \half C_{IJK} \e^{ab} b_a^J  \partial_\th b_b^K
\right) ,
\\
J_a &=& \frac{\partial \CF}{\partial \a^a} 
=
\frac{\pi}{4G_5}\int d\th \left[ 
\frac{\sqrt{G} w_2}{w_1^2} (G_{ab}\a^b + Q_{IJ} e^I b_a^J) 
-\frac{1}{3} C_{IJK} b^I_a \e^{bc} b^J_b \partial_\th b^K_c 
\right].
\label{jn2}
\ee
Finally, the entropy function is computed to be
\be
\CE &=& Q_I \tilde{e}^I + J_a \a^a - \CF 
\nn \\ 
&=& \frac{\pi}{4G_5}\int d\th \sqrt{G}  w_1^2 w_2 
\left( \frac{2}{w_1^2} +2V 
+ \frac{ G_{ab} \a^a\a^b}{2w_1^4} \right) 
+\frac{\pi}{4G_5} \left[ \frac{2}{w_2}\partial_\th \left(w_1^2\sqrt{G}\right)\right]_0^{\pi/2} 
\nn \\
&& - \frac{\pi}{4G_5}\int d\th \sqrt{G}  w_1^2 w_2 
\left( \frac{2\dot{w}_1^2}{w_1^2w_2^2} 
+\frac{4\dot{w}_1 \partial_\th \sqrt{G}}{w_1 w_2^2 \sqrt{G}}
+ \frac{1}{w_2^2}
\left(\frac{\partial_\th \sqrt{G}}{\sqrt{G}}\right)^2
+\frac{\partial_\th G^{ab} \partial_\th G_{ab}}{4w_2^2} 
\right)
\nn \\
&&+ \frac{\pi}{4G_5}\int d\th \sqrt{G}  w_1^2 w_2  \left( 
\frac{Q_{IJ} e^I e^J }{2w_1^4} 
+ \frac{Q_{IJ} G^{ab}  \partial_\th b_a^I \partial_\th b_b^J }{2w_2^2} 
+\frac{g_{ij} \partial_\th \vph^i \partial_\th \vph^j}{2w_2^2} 
\right).
\ee
Using the equations of motion, we can easily confirm the area law 
for the black hole entropy:
\be
2\pi \CE = \frac{\pi^2}{G_5} \int d\th \left( w_2 \sqrt{G} \right) + 
\frac{\pi^2}{2G_5} \left[ \frac{\sqrt{G} \partial_\th (w_1^2) }{w_2} \right]_0^{\pi/2}  
= \frac{({\rm Area})_{\rm hor}}{4G_5} = S_{\rm BH}.
\ee
The boundary term does not contribute to the entropy 
because both $\sqrt{G}$ and $\partial_\th(w_1^2)$ 
must vanish at $\th=0,\pi/2$ to avoid singular geometry.  
To reach this conclusion, it is essential to 
keep both contributions to the boundary term, 
one coming from (\ref{td1}) and the other from (\ref{td2}).

\subsection{Large black hole solutions}

To find a large black hole solution to the near horizon 
equations, we need to determine the dependence of 
the near horizon parameters on the size parameter $v$. 
For the non-derivative terms, 
we expect that the results from the equal rotation 
case continue to hold.:
\be
w_1 \sim 1, \;\; 
w_2 \sim v, \;\;
w_3 \sim v, \;\; 
e^I \sim 1, \;\; 
\a^a \sim v^{-2}, \;\;
b^I_a \sim v. 
\ee
The derivatives 
$\partial_\th w_2$, $\partial_\th w_3$ and $\partial_\th b_a^I$, 
are already present in the minimal supergravity solution, 
and we again assume the same dependence 
since all supergravity theories contain the minimal one: 
\be 
\partial_\th w_2 /w_2 \sim 1, \;\;
\partial_\th w_3 / w_3 \sim 1, \;\;
\partial_\th b_a^I \sim \; v.
\ee 
The remaining two terms, $\partial_\th w_1$ and $\partial_\th \vph^i$, 
are somewhat subtle. In minimal supergravity both of them 
are strictly zero, but in general they are expected to be non-zero. 
The two should be of the same order in $v$, 
as can be seen from, for example, 
\be
2 w_1^2 = - \frac{1}{V} =  Q_{IJ} e^I e^J  ,
\ee
which is obtained by taking the leading terms of the Einstein equations (\ref{einx1}, \ref{einx2}).
Very few explicit solutions with multiple charges and arbitrary rotations 
are known in the literature; 
refs. \cite{pope0} and \cite{reall1} are the only examples 
we are aware of. 
In these references, all the $\th$-dependences of $w_1$ and $\vph^i$ 
come in through the combination $\rho^2 = r^2 + a^2 c_\th^2+b^2s_\th^2$, 
which implies that, 
\be
\partial_\th w_1 \sim \partial_\th \vph^i \sim \frac{2  s_\th c_\th(b^2- a^2)}{v^2} .
\ee
We will proceed by assuming that this $1/v^2$ suppression is true of all 
large black hole solutions and later confirm that this 
assumption is self-consistent.

It is now straightforward to examine the near horizon equations 
to the order equations in $1/v$. 
The first two components of the Einstein equations 
as well as the scalar and Maxwell equations become quite simple. 
\be
2 w_1^2 = - \frac{1}{V} =  Q_{IJ} e^I e^J  ,
\\
\frac{\partial Q_{IJ}}{\partial \vph^i} \left( \frac{e^Ie^J}{w_1^4} \right)
- 4 \left(\frac{\partial V}{\partial \vph^i} \right) &=& 0,
\\
\frac{\sqrt{G}w_2}{w_1^2}  Q_{IJ} e^J \a^a  - C_{IJK} e^J \e^{ab} \partial_\th  b_b^K 
&=& 0,
\ee
The remaining equations (\ref{einx3}) and (\ref{einx4}) 
are also slightly simplified,
\be 
-\frac{\partial_\th}{w_2}  \left( \frac{\partial_\th\sqrt{G}}{w_2\sqrt{G}} \right)
+\frac{1}{w_2^2}\left(\frac{\partial_\th\sqrt{G}}{\sqrt{G}}\right)^2
+\frac{\partial_\th G^{ab} \partial_\th G_{ab}}{2w_2^2}
+\frac{ G_{ab} \a^a\a^b}{2w_1^4} - 
\frac{Q_{IJ} G^{ab}  \partial_\th b_a^I \partial_\th b_b^J }{2w_2^2} 
= 0.\;\;\;\;\;\;\;
\label{einy3}
\ee
\be
X_{ab} & -& \half (G^{cd}X_{cd})G_{ab} \; = \; 0,
\nn \\
X_{ab} &\equiv& \frac{1}{\sqrt{G}w_2} 
\left[   
\partial_\th\left( \frac{\sqrt{G}}{w_2} \partial_\th G_{ab} \right) 
- G_{ac} \partial_\th\left( \frac{\sqrt{G}}{w_2}  \partial_\th G^{cd} \right) G_{db} 
\right]
\nn \\
&&+ \frac{2(G_{ac}\a^c)(G_{bd}\a^d)}{w_1^4} 
+ \frac{2 Q_{IJ} \partial_\th b_a^I \partial_\th b_b^J}{w_2^2} .
\label{einy4}
\ee

The prediction from the fluid mechanics 
and the universal solution in the equal rotation case 
obtained earlier lead us to look for a universal solution 
even in the unequal rotation case. 
In fact, a straightforward computation shows that 
the following general solution satisfies all the 
leading order near horizon equations:
\be
\label{sol3met}
ds^2 &=& v_1^2 \left(-u^2dt^2 + \frac{du^2}{u^2} \right) 
+ \frac{v^2 d\th^2}{1-a^2 c_\th^2-b^2s_\th^2} 
+ v^2 G_{ab}(\th) D\phi^a D\phi^b,
\\
\label{sol3gau}
A^I &=& \tilde{e}^I udt + b^I_a(\th) D\phi^a  
\;\;\; 
\left( D\phi^a \equiv d\phi^a+\a^a u dt \right),
\ee
where
\be
(\a^1,\a^2) &=&\frac{4\sqrt{2} \bar{\a}v_1^2}{v^2} (b(1-a^2),a(1-b^2)),
\\
(b_1^I, b_2^I) &=& \sqrt{2} \bar{\a} \bar{b}^I v
\left( \frac{a s_\th^2}{1-a^2}, \frac{b c_\th^2}{1-b^2} \right) .
\ee
All the constants $(v_1, e^I, \bar{\a}, \bar{b}^I,\vph^i)$ are the same as 
in the previous section (\ref{new1}, \ref{new2}). 
The form of the metric, 
apart from the values of $\a^a$, 
has been taken from the minimal supergravity solution (\ref{sol2met2}).

\section{Discussions}

\subsection{Local entropy and charge densities}

The universal rotation dependence 
predicted by fluid mechanics \cite{minwa} determines not only 
the total entropy and charges but also their local density. 
Let us check how our computations on the horizon compare 
with the fluid mechanics on the boundary. 

{}From the solution (\ref{sol3met}), we find the entropy 
and  angular momentum densities
\be
 \frac{dS}{\sin\th \cos\th d\th} = \frac{S_0 
}{(1-a^2)(1-b^2)}, \;\;\;\;\;
\frac{d J_1}{\sin\th \cos\th d\th} = \frac{(a\sin^2\th) J_0}{(1-a^2)^2(1-b^2)}.
\ee
with $S_0$ and $J_0$ independent of $\th$. 
The charges $Q_I$ have the same angle dependence as $S$. 
The coordinates of \cite{pope1} we have been using 
is related to the manifestly asymptotically AdS coordinate by \cite{minwa}
\footnote{
The coordinate change is taken from eq. (141) of \cite{minwa}, 
which uses different names for the variables 
$(r, \th ; \tilde{r}, \chi)_{\rm here} = (y, \tilde{\th} ; r, \th)_{\rm there}$ and has a typographical error which we correct here. 
}
\be
\label{coch}
\tilde{r}^2\sin^2\chi = \frac{(r^2+a^2)\sin^2 \theta}{1-a^2}\, , 
\;\;\;\;\; 
\tilde{r}^2\cos^2\chi = \frac{(r^2+b^2)\cos^2 \theta}{1-b^2}\, . 
\ee
When $r,\tilde{r} \gg 1$, 
we can eliminate them to obtain 
the relation between the angles, 
\be
\cos^2\th = \frac{(1-b^2) \cos^2\chi}{(1-a^2 \sin^2\chi - b^2 \cos^2\chi)} .  
\ee
In terms of the $\chi$ coordinate, 
the local densities 
take the form, 
\be
\frac{dS}{\sin\chi \cos\chi d\chi} &=& 
\frac{S_0}{(1-a^2 \sin^2\chi -b^2 \cos^2\chi)^2} ,
\nn \\
\frac{d J_1}{\sin\chi \cos\chi d\chi} &=& 
\frac{(a \sin^2\chi) J_0}{(1-a^2 \sin^2\chi -b^2 \cos^2\chi)^3} ,
\ee
which coincides precisely with the expressions 
obtained from the fluid mechanics in \cite{minwa}. 

Note that this agreement is not quite trivial. 
The fluid mechanics computation was performed over 
the round $S^3$ at the boundary of the AdS, 
while our computation was done on the black hole horizon 
which has the shape of a squashed $S^3$ due to the rotations. 
In general, the map between a region of the fluid 
and the corresponding region on the horizon could be quite complicated. 
Our result above shows that such a complication does not occur 
to the leading order in the large black hole limit.

\subsection{The $h$ function}

The entropy function formalism enables us to express the 
entropy $S$, angular momenta $J_a$ and electric charges $Q_I$ 
in terms of the near horizon data without ever referring to 
the full solution. 
In contrast, the mass $E$, angular velocity $\w_a$ and the 
electric potential $\m^I$ cannot be computed in general 
from the near horizon data only. 

In the large black hole limit, however, the simple 
mapping to the fluid mechanics we have discussed so far 
enables us to write down $(E,\O^a, \m^I)$ in terms of the 
near horizon data. First, by comparing 
\be
S &=& \left(\frac{\pi}{4G}\right) \frac{2\pi v^3}{(1-a^2)(1-b^2)},
\\
\label{muu2}
Q_I &=& \left(\frac{\pi}{4G}\right) \frac{v^3Q_{IJ} e^J}{2v_1^2(1-a^2)(1-b^2)}  ,  
\\
J_1 &=&  \left(\frac{\pi}{4G}\right)
\frac{\sqrt{2} v^4 Q_{IJ} e^I \bar{b}^J \bar{\a}}{8v_1^2 (1-a^2)(1-b^2)} 
\left[ \frac{2a}{1-a^2} \right] , 
\ee
with the predictions of fluid mechanics, 
we note the identification, 
\be
\O_1 = a, \;\;\; \O_2= b ,
\ee
and the $T\goto 0$ limit of the $h$ function,  
\be
T^4 h \goto 
\frac{\sqrt{2}v^4 Q_{IJ} e^I \bar{b}^J \bar{\a}}{ 64 \pi G v_1^2}, 
\;\;\;
 T^3 \p_I h  \goto
\frac{v^3 Q_{IJ} e^J}{16\pi G v_1^2} ,
\;\;\;
T^3 \left(4h - \n^I \p_I h \right) \goto
\frac{2\pi v^3}{8\pi G} .
\ee
Next, since both $J_a$ and $E$ are proportional to the function $h$ 
with a universal rotation dependence, we have 
\be
E = 
\left(\frac{\pi}{4G}\right) 
\frac{\sqrt{2} v^4 Q_{IJ} e^I \bar{b}^J \bar{\a}}{8v_1^2 (1-a^2)(1-b^2)} 
\left[ \frac{2a^2}{1-a^2}+\frac{2b^2}{1-b^2}+3 \right] .
\ee 
Note the estimate for the mean free path
\be
l_{\rm mfp} \sim \left. \frac{S}{4\pi E} \right|_{\O=0} 
\sim \frac{1}{v}\left(\frac{v_1^2}{Q_{IJ}e^I\bar{b}^J \bar{\a}}\right) , 
\ee
where the quantity in the parenthesis is generically of order one. 
So we confirm that the fluid dynamics is indeed a good description 
in the limit of large $v$.

Finally, the chemical potentials are obtained by noting that, in the absence of rotation, 
$
E = \frac{3}{4} \mu^I Q_I 
$
should hold (See section 2.5 of \cite{minwa}):
\be
\mu^I  =  \sqrt{2} v \bar{\a} \bar{b}^I .
\ee


All gauged supergravity theories contain the minimal supergravity 
as a closed sub-sector. The minimal supergravity admit large 
extremal black hole solutions. It is reasonable to expect that, 
generically, small deformation away from the minimal supergravity 
will not change the properties of the black hole drastically, 
as we confirmed in the two examples above. 
Thus, large extremal black holes are rather generic objects in 
gauged supergravity.  

{}From the CFT point of view, this means that the zero temperature 
limit of the CFT ``fluid" is smooth; all physical quantities 
$(S,J_a,Q_I;E,\O_a,\m^I)$ remain finite as $T$ approaches zero. 
In order for $E$ and $J_a$  to have a smooth limit as $\n_I=\m_I/T$ become large, 
the $h(\n)$ function should approach a homogeneous function 
of degree four in the leading order:
\be
h(\n) = h_4(\n) + \mbox{(sub-leading)} , \;\;\; h_4(\l\n) = \l^4 h(\n).
\ee
Recall that the entropy $S$ is proportional to $T^3(4h-\n_I\p_Ih)$. 
The leading term $h_4(\n)$ does not contribute to $S$. 
The only possible contribution to $S$ could come from a homogeneous function 
of degree three, 
\be
h(\n) = h_4(\n) + h_3(\n) + \mbox{(sub-leading)} , \;\;\; 
h_k(\l\n) = \l^k h(\n).
\ee

In the $U(1)^3$ theory, up to an overall normalization, 
the $h$ function and its ``descendants" are given by \cite{minwa}
\be
h = \frac{L_1L_2L_3}{(L_1+L_2+L_3-1)^4} ,
&&
\n_I = \frac{\sqrt{L_I(1-L_I)}}{L_1+L_2+L_3-1} ,
\\
\p_I h = \frac{2L_1L_2L_3}{(L_1+L_2+L_3-1)^3} \sqrt{\frac{1-L_I}{L_I}} ,
&& 4h- \n^I \p_I h = \frac{2L_1L_2L_3}{(L_1+L_2+L_3-1)^3} .
\ee
In the $T\goto 0$ limit, the three variables $L_I$ approaches the 
``extremal triangle'' ($\sum_I L_I=1, L_I>0$). 
In this limit, the leading behavior of the $h$-function is manifestly 
quartic in $\n$ 
and it is straightforward to separate $h_4(\n)$ explicitly:
\be
h_4(\n) &=& \frac{1}{4}\left[2(\n_1^2 \n_2^2+\n_2^2 \n_3^2+\n_3^2 \n_1^2)-(\n_1^4+\n_2^4+\n_3^4)\right] 
\\
&=& \frac{1}{4}(\n_1+\n_2+\n_3)(-\n_1+\n_2+\n_3)(\n_1-\n_2+\n_3)(\n_1+\n_2-\n_3).
\ee
It is also not difficult to extract $h_3(\n)$:
\be
h_3(\n) = \sqrt{(-\n_1^2+\n_2^2+\n_3^2)(\n_1^2-\n_2^2+\n_3^2)(\n_1^2+\n_2^2-\n_3^2)/2} .
\ee

It would be interesting to understand 
further the functions $h_4$ and $h_3$ 
from both the supergravity side and the dual CFT side.
{}From the supergravity point of view, these functions can 
depend only on the parameters of the Lagrangian, namely, 
$C_{IJK}$ and $\bar{X}_I$.
It follows that, if $h_4$ is a polynomial, then the coefficients 
in the expansion, 
\be
h_4(\n) = \frac{1}{4!} h_{IJKL} \n^I\n^J\n^K\n^L ,
\ee
should be constants composed of $C_{IJK}$ and $\bar{X}_I$. 
Unfortunately, it appears that $h_4$ is not a polynomial in general.
We have checked whether the $h_4$ of the conifold CFT example 
discussed in section 3.2.4 can be expressed as a polynomial in $\n^I$ 
and found the answer in the negative. 

\subsection{Future directions}

The near horizon analysis performed in this paper 
should be regarded as a first step to reveal the 
connection between the two holographic screens, 
namely, the black hole horizon and the AdS boundary. 
There are several directions one may pursue further. 

We restricted our attention to the extremal black holes only. 
It will be clearly useful to do similar computations 
for black holes with non-zero temperature 
and make a more comprehensive comparison with the fluid dynamics.  
Non-extremal entropy function of \cite{cai} may be useful in this regard. 

The entropy function we obtained could 
be used away from the large black hole limit. 
For instance, it could be used to explore the existence of 
supersymmetric black holes in theories 
where no solutions have been constructed. 

Finally, it would be interesting to compute the subleading corrections 
both in the fluid mechanics and on the near horizon equations 
and see how the map between the two changes. 
Recently, some progress in this direction was reported in \cite{minwa2}, 
where it was shown, to the first subleading order,  
how the region of the fluid evolves in the radial direction 
in the case of uncharged black branes. 
Incorporating electric charges and working in the global AdS 
rather than the Poincar\'e patch, the corrections in the fluid mechanics 
could be matched against corresponding corrections 
to the near horizon equations we obtained in this paper. 
The systematic derivative expansion developed in 
\cite{minwa2, bai, nat2, nat3, loga} would be helpful in such an attempt.
We hope to return to some of these questions in the near future.

\vskip 1.5cm

\subsection*{Acknowledgment}

It is our pleasure to thank 
K. Goldstein, S. Trivedi and H. Yavartanoo 
and especially S. Minwalla 
for useful discussions and/or correspondences, 
We also thank Seok Kim for collaboration at 
early stages of this work. 
Sangmin Lee is grateful to 
the organizers of the 2nd Asian School on String Theory 
where part of the work was done, 
and 
the String Theory Group 
and the Institute of Mathematical Sciences 
at Imperial College London for hospitality 
during his visit at the final stage of the work.
The work of Sangmin Lee is supported in part by the KOSEF Grant R01-2006-000-10965-0 and the 
Korea Research Foundation Grant KRF-2007-331-C00073. 
The work of Sungjay Lee is supported in part by the Korea
Research Foundation Grant R14-2003-012-01001-0.

\vskip 0.8cm

\centerline{\large \bf Appendix}

\appendix

\section{Chern-Simons contribution to the entropy function}

A very general and thorough discussion of how to deal with Chern-Simons 
contributions to the entropy function can be found in \cite{sen5}. 
In our case, in essence, the correct procedure amounts to 
maintaing the radial dependence of the variables $b^I$ in 
the near horizon ansatz,
\be
\label{ansatzx}
A^I &=&  \tilde{e}^I u dt + b^I (\s_3 + \a u dt) \; \equiv \; 
\tilde{e}^I u dt + b^I \tilde{\s}_3,
\nn \\ 
F^I &=& (\tilde{e}^I +\a b^I) du \wedge dt - b^I \s_1 \wedge \s_2 
+ (\partial_u b^I) \, du \wedge \tilde{\s}_3 \; ,
\ee
in the intermediate steps and setting $b^I$ to constant 
only at the final step. 

Inserting (\ref{ansatzx}) into 
the Chern-Simons term of the Lagrangian, 
\be
(16\pi G)\CL_{\rm CS} = -\frac{1}{6} C_{IJK} F^I \wedge F^J \wedge A^K ,
\ee
we find 
\begin{eqnarray}
  \left(16 \pi G\right) \CL_{\rm CS} &=& \left( 16 \pi G \right) \left( \CL_{\rm CS}^1 + \CL_{\rm CS}^2 \right) \, , 
\\
(16\pi G) \CL_{\rm CS}^1 &=&
\frac{1}{3} C_{IJK}
(b^I \tilde{\s}_3) ((\tilde{e}^J+\a b^J) du dt) (b^K \s_1 \s_2)
\nn \\
&=& - \frac{1}{3} C_{IJK}( \tilde{e}^I b^J b^K + \a b^I b^J b^K)
(dt du \s_1 \s_2 \tilde{\s}_3)\ ,
\\
\label{cs2}
(16\pi G) \CL_{\rm CS}^2 &=&
\frac{1}{3} C_{IJK}
(\tilde{e}^I u dt) (\partial_u b^J du \tilde{\s}_3) (b^K \s_1 \s_2)
\nn \\
&=&
\frac{1}{6} C_{IJK} \tilde{e}^I u \partial_u (b^J b^K)
(dt du \s_1 \s_2 \tilde{\s}_3)
\nn \\
&=&
- \frac{1}{6} C_{IJK} \tilde{e}^I b^J b^K
(dt du \s_1 \s_2 \tilde{\s}_3) + \mbox{(total derivative)} .
\ee
Adding the two terms and discarding the total derivative term, 
we obtain
\begin{eqnarray}
\label{csf}
  \left(16 \pi G\right) \CL_{\rm CS} 
    = - C_{IJK} \left( \frac12  \tilde{e}^I b^J b^K + \frac13 \alpha b^I b^J b^K \right) \left( dt du \s_1 \s_2 \tilde{\s}_3 \right)
  \, .
\end{eqnarray}
A naive evaluation of the near horizon action 
would fail to include the second contribution (\ref{cs2}). 
An alternative way to arrive at the correct answer (\ref{csf}) 
is to take a dimensional reduction along the $\s_3$ direction 
to make the problem effectively four dimensional \cite{kevin}.




\end{document}